\documentclass{article}

\usepackage{arxiv}

\usepackage[utf8]{inputenc} %
\usepackage[T1]{fontenc}    %
\usepackage{hyperref}       %
\usepackage{url}            %
\usepackage{booktabs}       %
\usepackage{amsfonts}       %
\usepackage{nicefrac}       %
\usepackage{microtype}      %
\usepackage{lipsum}
\usepackage{graphicx}
\usepackage{caption}
\usepackage{subcaption}
\usepackage{color}
\usepackage{amsmath}
\usepackage[numbers]{natbib}
\graphicspath{ {./images/} }

\usepackage{lineno}

\usepackage{setspace}
\title{Asynchronous Deep Double Duelling Q-Learning for Trading-Signal Execution in Limit Order Book Markets}

\author{
 Peer Nagy \\
  Oxford-Man Institute of Quantitative Finance\\
  Department of Engineering Science\\
  University of Oxford\\
  \texttt{peer.nagy@eng.ox.ac.uk} \\
   \And
 Jan-Peter Calliess \\
  Oxford-Man Institute of Quantitative Finance\\
  Department of Engineering Science\\
  University of Oxford\\
  \And
 Stefan Zohren \\
  Oxford-Man Institute of Quantitative Finance\\
  Department of Engineering Science\\
  University of Oxford\\
}

\begin{document}

\maketitle
\begin{abstract}
We employ deep reinforcement learning (RL) to train an agent to successfully translate a high-frequency trading signal into a trading strategy that places individual limit orders.
Based on the ABIDES limit order book simulator, we build a reinforcement learning OpenAI gym environment and utilise it to simulate a realistic trading environment for NASDAQ equities based on historic order book messages.
To train a trading agent that learns to maximise its trading return in this environment, we use Deep Duelling Double Q-learning with the APEX (asynchronous prioritised experience replay) architecture. The agent observes the current limit order book state, its recent history, and a short-term directional forecast. To investigate the performance of RL for adaptive trading independently from a concrete forecasting algorithm, we study the performance of our approach utilising synthetic alpha signals obtained by perturbing forward-looking returns with varying levels of noise. Here, we find that the RL agent learns an effective trading strategy for inventory management and order placing that outperforms a heuristic benchmark trading strategy having access to the same signal.
\end{abstract}


\section{Introduction}

Successful quantitative trading strategies often work by generating trading signals, which exhibit a statistically significant correlation with future prices. These signals are then turned into actions, aiming to assume positions in order to gain from future price changes. The higher the signal frequency and strategy turnover, the more critical is the execution component of the strategy, which translates the signal into concrete orders that can be submitted to a market. Such markets are oftentimes organised as an order ledger represented by a \emph{limit order book (LOB)}.

Limit order book prices have been shown to be predictable over short time periods, predicting a few successive ticks into the future with some accuracy.
This has been done by either utilising the recent history of order book states \cite{deeplob_zhang19, multi_horizon_lob_zhang21}, order-flow data \cite{deep_ofi_kolm21}, or market-by-order (MBO) data directly as features \cite{deep_mbo_zhang21}.
However, given the short time horizons over which these predictions are formed, and correspondingly small price movements, predictability does not directly translate into trading profits.
Transaction costs, strategy implementation details, and time delays add up to the challenging problem of translating high-frequency forecasts into a trading strategy that determines when and which orders to send to the exchange. In addition, different predictive signals have to be traded differently to achieve optimal results, depending on the forecast horizon, signal stability, and predictive power.

In this paper, we use asynchronous off-policy reinforcement learning (RL), specifically \emph{Deep Duelling Double Q-learning} with the \emph{APEX} architecture \cite{apex_horgan18}, to learn an optimal trading strategy, given a noisy directional signal of short-term forward mid-quote returns. For this purpose, we developed an OpenAI gym \cite{gym_brockman16} limit order book environment based on the ABIDES \cite{abides_byrd20} market simulator, similar to \cite{abides_gym_amrouni21}. We use this simulator to replay NASDAQ price-time priority limit order book markets using message data from the LOBSTER data set \cite{lobster_huang11}.

If a financial trader wants to transact a number of shares of a financial security, such as shares of cash equities, for example Apple stock, they need to send an order to an exchange. Most stock exchanges today accept orders electronically and in 2022 in US equity markets approximately 60 - 73\% of orders are submitted by algorithms, not human traders \cite{mordor}.
A common market mechanism to efficiently transact shares between buyers and sellers is the \emph{limit order book (LOB)}. The LOB contains all open limit orders for a given security and is endowed with a set of rules to clear marketable orders. A more detailed introduction to LOBs can be found in section \ref{subsec:lob}.

High-frequency trading (HFT) is an industry with high barriers to entry and a regulatory landscape varying by geographical region. Generally, trading on both sides of the LOB simultaneously by submitting both limit buy and limit sell orders is the purview of \emph{market makers}. While their principal role is to provide liquidity to the market, they also frequently take directional bets over short time periods. Transaction costs for such trading strategies consist of two main components, \emph{explicit costs}, such as trading fees, and \emph{implicit costs}, such as market impact \cite{harris2003trading}. Trading fees vary by institution but can be negligible for institutional market makers, with some fee structures even resulting in zero additional costs if positions are closed out at the end of the day. The main consideration of transaction costs should thus be given to market impact. Our simulation environment models direct market impact by injecting new orders into historic order flow, thereby adding to or consuming liquidity in the market. One limitation of this approach is that indirect market impact and, generally, the reactions of other market participants are not modelled.

We study the case of an artificial or synthetic signal, taking the future price as known and adding varying levels of noise, allowing us to investigate learning performance and to quantify the benefit of an RL-derived trading policy compared to a baseline strategy using the same noisy signal. This is not an unrealistic setup when choosing the correct level of noise. Practitioners often have dedicated teams researching and deriving alpha signals, often over many years, while other teams might work on translating those signals into profitable strategies. Our aim is to focus on the latter problem, which becomes increasingly more difficult as signals become faster. It is thus interesting to see how an RL framework can be used to solve this problem. In particular, we show that the RL agent learns policies superior to the baselines, both in terms of strategy return and Sharpe ratio. Machine learning methods, such as RL, have become increasingly important to automate trade execution in the financial industry in recent years \cite{cfa_ml_execution_nagy23}, underlining the practical use of research in this area.

We make a number of contributions to the existing literature. By defining a novel action and state space in a LOB trading environment, we allow for the placement of limit orders at different prices. This allows the agent to learn a concrete high-frequency trading strategy for a given signal, trading either aggressively by crossing the spread, or conservatively, implicitly trading off execution probability and cost. In addition to the timing and level placement of limit orders, our RL agent also learns to use limit orders of single units of stock to manage its inventory as it holds variably sized long or short positions over time. 
More broadly, we demonstrate the practical use case of RL to translate predictive signals into limit order trading strategies, which is still usually a hand-crafted component of a trading system. We thus show that simulating limit order book markets and using RL to further automate the investment process is a promising direction for further research. To the best of our knowledge, this is also the first study applying the APEX \cite{apex_horgan18} algorithm to limit order book environments.

The remaining paper is structured as follows:
Section \ref{s:related_work} surveys related literature, section \ref{s:background} explains the mechanics of limit order book markets and the APEX algorithm,
section \ref{s:framework} details the construction of the artificial price signal, section \ref{s:results} showcases our empirical results,
and section \ref{s:conclusion} concludes our findings.

\section{Related Work} \label{s:related_work}

Reinforcement learning has been applied to learn different tasks in limit order book market environments, such as optimal trade execution \cite{nevmyvaka06, ddqn_opt_ex_ning18, daberius19, karpe20, schnaubelt22}, market making \cite{mm_abernethy13, drl_mm_kumar20}, portfolio optimisation \cite{mb_rl_portf_opt_yu19} or trading \cite{kearns13, model_based_rl_lob_wei19, briola21}. 
The objective of optimal trade execution is to minimise the cost of trading a predetermined amount of shares over a given time frame. Trading direction and the number of shares is already pre-defined in the execution problem. Market makers, on the other hand, place limit orders on both sides of the order book and set out to maximise profits from capturing the spread, while minimising the risk of inventory accumulation and adverse selection. We summarise using the term ``RL for trading'' such tasks which maximise profit from taking directional bets in the market. This is a hard problem for RL to solve as the space of potential trading strategies is large, leading to potentially many local optima in the loss landscape, and actionable directional market forecasts are notoriously difficult due to arbitrage in the market.

The work of \cite{kearns13} is an early study of RL for market microstructure tasks, including trade execution and predicting price movements. While the authors achieve some predictive power of directional price moves, forecasts are determined to be too erroneous for profitable trading. The most similar work to ours is \cite{briola21} that provides the first end-to-end DRL framework for high-frequency trading, using PPO \cite{ppo_schulman17} to trade Intel stock. To model price impact, \cite{briola21} use an approximation, moving prices proportionately to the square-root of traded volume. The action space is essentially limited to market orders, so there is no decision made on limit prices. The trained policy is capable of producing a profitable trading strategy on the evaluated 20 test days. However, this is not compared to baseline strategies and the resulting performance is not statistically tested for significance. 
In contrast, we consider a larger action space, allowing for the placement of limit orders at different prices, thereby potentially lowering transaction costs of the learned HFT strategy.
For a broader survey of deep RL (DRL) for trading, including portfolio optimisation, model-based and hierarchical RL approaches the reader is referred to \cite{drl_mm_kumar20}.

One strand of the literature formulates trading strategies in order-driven markets, such as LOBs, as optimal stochastic control problems, which can often even be solved analytically. A seminal work in this tradition is \citet{almgren_chriss01}, which solves a simple optimal execution problem. More recently, \citet{cartea2018enhancing} use order book imbalance, i.e. the relative difference in volume between buy and sell limit orders, to forecast the direction of subsequent market orders and price moves. They find that utilising this metric can thereby improve the performance of trading strategies. In addition to market orders, LOB imbalance has also been found to be predictive of limit order arrivals, and illegal manipulation of the LOB imbalance by \emph{spoofing} can be a profitable strategy \cite{cartea2020spoofing}. Stochastic models, assuming temporary and permanent price impact functions have also found that using order flow information can reduce trading costs when trading multiple assets \cite{cartea2019trading, cartea2016incorporating}.
In contrast to the stochastic modelling literature, we employ a purely data-driven approach using simulation. This allows us to make fewer assumptions about market dynamics, such as specific functional forms and model parameters. The stochasticity of the market is captured in large samples of concrete data realisations.

\section{Background} \label{s:background}
\subsection{Limit Order Book Data} \label{subsec:lob}

Limit order books (LOBs) are one of the most popular financial market mechanisms used by exchanges around the world \cite{lob_gould13}. Market participants submit limit buy or sell orders, specifying a maximum (minimum) price at which they are willing to buy (sell), and the size of the order. The exchange's LOB then keeps track of unfilled limit orders on the buy side (bids) and the sell side (asks). If an incoming order is marketable, i.e. there are open orders on the opposing side of the order book at acceptable prices, the order is matched immediately, thereby removing liquidity from the book. The most popular matching prioritisation scheme is price-time priority. Here, limit orders are matched first based on price, starting with the most favourable price for the incoming order, and then based on arrival time, starting with the oldest resting limit order in the book, at each price level.

\begin{figure*}[h!]
\centering
\includegraphics[width=0.5\textwidth]{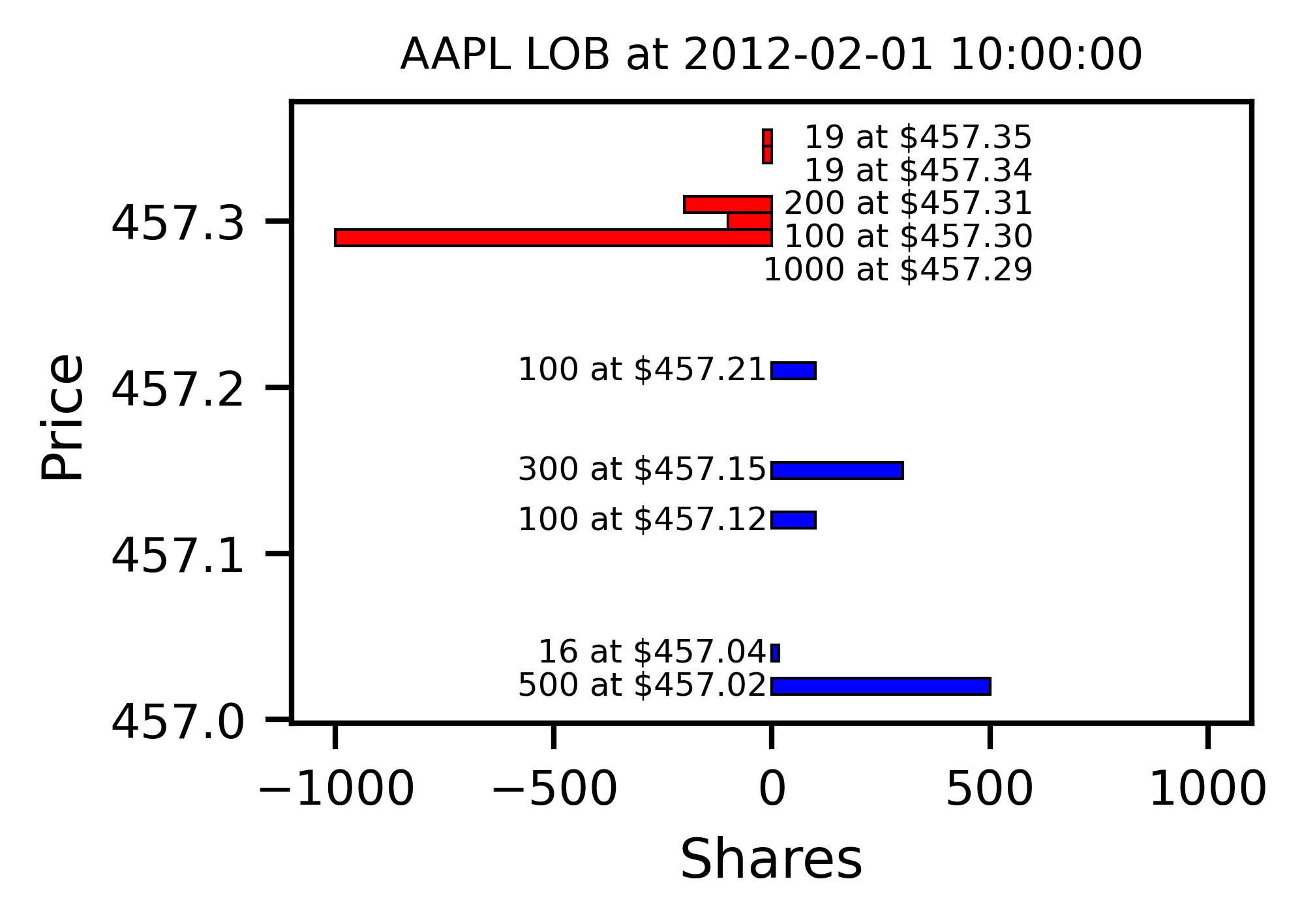}
\caption{Example Limit Order Book Snapshot of Apple Stock (AAPL) on 01 February 2012 at 10 am. Displayed are the 5 best bid (blue) and ask  levels (red). Ask sizes are shown as negative values to indicate limit \emph{sell} orders. AAPL is an example of a small-tick stock since the minimum tick size of \textcent1 is small compared to the stock price. For a given number of orders, this results in sparser books with more empty price levels. (NASDAQ data from LOBSTER \cite{lobster_huang11})}
\label{fig:aapl_lob}
\end{figure*}

Figure \ref{fig:aapl_lob} shows an example snapshot of an Apple LOB, traded at the NASDAQ exchange, on Wednesday 01 February 2012 at 10 am. Shown are the best 5 price levels on the bid and ask side and the aggregated available volume at each price. In this example, the best bid lies at \$457.21 for 100 shares, which is the maximum instantaneous price a potential seller would be able to trade at, for example using a market order. Conversely, a potential buyer could receive up to 1000 shares at a price of \$457.29. Trading larger quantities on either side would use up all volume at the best price and consume additional liquidity at deeper levels. Submitting a buy limit order at a price below the best ask would not be marketable immediately and instead enter the LOB as new volume on the bid side.
For a more comprehensive review of limit order book dynamics and pertaining models, we refer the reader to \cite{lob_gould13}.

In this paper, we consider equity limit order book data from the NASDAQ exchange \cite{lobster_huang11}, which also uses a price-time priority prioritisation. Our market simulator keeps track of the state of the LOB by replaying historical message data, consisting of new incoming limit orders, order cancellations, or modifications. The RL agent can then inject new messages into the order flow and thereby, change the LOB state from its observed historical state.

Our simulator reconstructs LOB dynamics from message data, so every marketable order takes liquidity from the book and thus has a direct price impact. Beyond that, we make no further assumptions on permanent market impact or reactions of other agents in the market, which we leave to future work.

\subsection{Deep Reinforcement Learning}

We model the trader's problem as a Markov Decision Process (MDP) \cite{mdp_bellman57, mdp_puterman90}, described by the tuple $\left<\mathcal{S}, \mathcal{A}, \mathcal{T}, r, \gamma\right>$. $\mathcal{S}$ denotes a state space, $\mathcal{A}$ an action space, $\mathcal T$ a stochastic transition function, $r$ a reward function, and $\gamma$ a discount factor. 
Observing the current environment state $s_t \in \mathcal{S}$ at time $t$, the trader takes action $a_t \in \mathcal{A}$, which causes the environment to transition state according to the stochastic transition function $\mathcal{T}(s_{t+1}|s_t, a_t)$. After transitioning from $s_t$ to $s_{t+1}$, the agent receives a reward $r_{t+1} = r(s_t,a_t,s_{t+1})$.

Solving the MDP amounts to finding an optimal policy $\pi^*: \mathcal{S} \to \mathcal{A}$, which maximises the discounted expected sum of future rewards $\sum_{i=t+1}^{T} \gamma^{i-t} \mathbb{E}_t r_{i}$ between current time $t$ and terminal time $T$, given a discount factor $\gamma \in (0,1]$. As the transition kernel $\mathcal{T}$ is unknown, we use reinforcement learning (RL) to learn an optimal policy from observed trajectories of state-action transitions. RL algorithms fall broadly within two categories: value-based methods, which learn representations of value functions, and policy-based methods, which learn explicit representations of the agent's policy. \footnote{Actor-critic algorithms fall between the two as they keep explicit representations of both policy (actor) and value functions (critic).}
In this paper, we are using a value-based RL algorithm, based on \emph{Q-learning}, which explicitly approximates the action-value function $Q^*$ for the optimal policy $\pi^*$. The action-value function for a given policy $\pi$ is defined recursively as $Q^{\pi}(s, a) = \mathbb{E}_t r_{t+1} + \gamma \max_{a'} Q^{\pi}(s_{t+1}, a')$.
One benefit of this class of algorithms is that they are \emph{off-policy} learners, which means that they approximate the optimal value function $Q^*$ using transitions sampled from the environment using a, potentially sub-optimal, exploratory policy $\pi$. This allows for computational efficiency due to asynchronous sampling and learning steps, as described in the next section.
Modern Deep RL algorithms, such as DQN \cite{dqn_mnih13}, use neural networks as function approximators, in this case, of the Q-function. This way algorithms make use of the generalisation abilities provided by deep learning, necessitated by large or continuous state spaces. Network parameters are updated using \emph{temporal difference learning} with gradient-based optimisers, such as stochastic gradient descent or the popular Adam optimizer \cite{kingma2014adam}.
For a comprehensive treatment of RL, we refer the interested reader to \citet{sutton2018reinforcement}, and for a survey of recent progress in RL for financial applications to \citet{hambly2023recent}.

\subsection{Double DQN with Distributed Experience Replay}

We use Deep Double Q-learning \cite{dddqn_hasselt2016} with a duelling network architecture \cite{duelling_wang16} to approximate the optimal Q-function $Q^*(s,a)= \mathbb{E}[r_{t+1} + \gamma \max_{a'} Q^*(s_{t+1},a') | a_t=a, s_t=s]$. To speed up the learning process we employ the \emph{APEX} training architecture \cite{apex_horgan18}, which combines asynchronous experience sampling using parallel environments with off-policy learning from experience replay buffers. Every episode $i$ results in an experience trajectory $\tau_i=\{s_t,a_t\}_{t=1}^T$, many of which are sampled from parallel environment instances and are then stored in the replay buffer. The environment sampling is done asynchronously using parallel processes running on CPUs. Experience data from the buffer is then sampled randomly and batched to perform a policy improvement step of the Q-network on the GPU. Prioritised sampling from the experience buffer has proven to degrade performance in our noisy problem setting, hence we are sampling uniformly from the buffer.\footnote{In many application domains prioritised sampling, whereby we resample instances more frequently where the model initially performs poorly tends to aide learning. However, in our low signal-to-noise application domain, we noted poor performance. Investigating the matter, we found that prioritised sampling caused more frequent resampling of highly noisy instances where learning was particularly difficult, thus degrading performance.} After a sufficient number of training steps, the new policy is then copied to every CPU worker to update the behavioural policy.

Double Q-learning \cite{ddqn_hasselt10, dddqn_hasselt2016} stabilises the learning process by keeping separate Q-network weights for action selection (main network) and action validation (target network). The target network weights are then updated gradually in the direction of the main network's weight every few iterations. Classical Q-learning without a separate target network can be unstable due to a positive bias introduced by the $\max$ operator in the definition of the Q-function, leading to exploding Q-values during training.
The duelling network architecture \cite{duelling_wang16} additionally uses two separate network branches (for both main and target Q-networks). One branch estimates the value function $V(s)=\max_a Q(s, a)$, while the other estimates the advantage function $A(s, a) = Q(s, a) - V(s)$. The benefit of this architecture choice lies therein that the advantage of individual actions in some states might be irrelevant, and the state value, which can be learnt more easily, suffices for an action-value approximation. This leads to faster policy convergence during training.

\section{Framework} \label{s:framework}

\subsection{Artificial Price Signal}

The artificial directional price signal $d_t \in \Delta^2 = \{x \in \mathbb{R}^3: x_1 + x_2 + x_3 = 1, \; x_i \geq 0 \;\text{for}\; i = 1,2,3\}$ the agent receives is modelled as a discrete probability distribution over 3 classes,
corresponding to the averaged mid-quote price decreasing, remaining stable, or increasing over a fixed future time horizon of $h \in \mathbb{N}_+$ seconds.
To achieve realistic levels of temporal stability of the signal process, $d_t$ is an exponentially weighted average, with persistence coefficient $\phi \in (0,1)$, of Dirichlet random variables $\epsilon_t$. The Dirichlet parameters $\alpha$ depend on the realised smoothed future return $\overline{r_{t+h}}$, specifically on whether the return lies within a neighbourhood of size $k$ around zero, or above or below. Thus we have:

\begin{equation}
\begin{aligned}
    d_t &= \phi d_{t-1} + (1-\phi) \epsilon_t \\
    \epsilon_{t} &= \textrm{Dirichlet}\left(
        \alpha(\overline{r_{t+h}})
    \right)\\
    \overline{r_{t+h}} = \frac{\overline{p_{t+h}} - p_t}{p_t} &\quad \text{where} \quad \overline{p_{t+h}} = \frac{1}{h} \sum_{i=1}^{h} p_{t+i}
\end{aligned}
\end{equation} \label{equ:oracle_signal}

and prices $p_t$ refer to the mid-quote price at time $t$. The Dirichlet distribution is parametrised, so that, in expectation, the signal $d_t$  updates in the direction of future returns, where $a^H$ and $a^L$ determine the variance of the signal. The Dirichlet parameter vector is thus:

\begin{equation}
\alpha(r_{t+h}) =
    \begin{cases}
        (a^H, a^L, a^L)  & \text{if } \overline{r_{t+h}} < -k \\
        (a^L, a^H, a^L)  & \text{if } -k \leq \overline{r_{t+h}} < k \\
        (a^L, a^L, a^H)  & \text{if } k \leq \overline{r_{t+h}}. \\
    \end{cases}
\end{equation}

\subsection{RL Problem Specification} \label{subsec:rl_spec}

At each time step $t$, the agent receives a new state observation $s_t$. $s_t$ consists of the time left in the current episode $T-t$ given the episode's duration of $T$, the agent's cash balance $C_t$, stock inventory $X_t$, the directional signal $d_t \in \Delta^2$, encoded as probabilities of prices decreasing, remaining approximately constant, or increasing; and price and volume quantities for the best bid and ask (level 1), including the agent's own volume posted at bid and ask: $o_{b,t}$ and $o_{a, t}$ respectively.
In addition to the most current observable variables at time t, the agent also observes a history of the previous $l$ values, which are updated whenever there is an observed change in the LOB. Putting all this together, we obtain the following state observation:

$$
s_t = \begin{pmatrix}
        T-u\\% \tau_u\\
        C_u\\
        X_u\\
        (d_u^1, d_u^2, d_u^3)'\\
        (p_{a, u}, v_{a, u}, o_{a, u}, p_{b, u}, v_{b, u}, o_{b, u})'
       \end{pmatrix}_{u=\{t-l,\ldots,t\}}.
$$

After receiving the state observation, the agent then chooses an action $a_t$. It can place a buy or sell limit order of a single share at bid, mid-quote, or ask price; or do nothing and advance to the next time step. Actions, which would immediately result in positions outside the allowed inventory constraints $[pos_{min}, pos_{max}]$ are disallowed and do not trigger an order. Whenever the execution of a resting limit order takes the inventory outside the allowed constraints, a market order in the opposing direction is triggered to reduce the position back to $pos_{min}$ for short positions or $pos_{max}$ for long positions. Hence, we define

$$
a_t \in \mathcal{A} = \left( \{-1, 1\} \times \{-1,0,1\} \right) \cup \{skip\}
$$
so that in total there are 7 discrete actions available, three levels for both buy and sell orders, and a skip action. For the six actions besides the ``skip'' action, the first dimension encodes the trading direction (sell or buy) and the second dimension the price level (bid, mid-price, or ask). For example, $a=(1, 0)$ describes the action to place a buy order at the mid price, and $a=(-1,1)$ a sell order at best ask. Rewards $R_{t+1}$ consist of a convex combination of a profit-and-loss-based reward $R_{t+1}^{pnl}$ and a directional reward $R_{t+1}^{dir}$. $R_{t+1}^{pnl}$ is the log return of the agent's mark-to-market portfolio value $M_t$, encompassing cash and the current inventory value, marked at the mid-price. The benefit of log-returns is that they are additive over time, rather than multiplicative like gross returns, so that, without discounting ($\gamma=1$) the total profit-and-loss return $\sum_{s=t+1}^T R_s^{pnl} = M_T - M_t$.
The directional reward term $R_{t+1}^{dir}$ incentivizes the agent to hold inventory in the direction of the signal and penalises the agent for inventory positions opposing the signal direction. The size of the directional reward can be scaled by the parameter $\kappa > 0$. $R_{t+1}^{dir}$ is positive if the positive prediction has a higher score than the negative ($d_{t,3} > d_{t,1}$) and the current inventory is positive; or if $d_{t,3} < d_{t,1}$ and $X_t < 0$. Further, if the signal $[-1, 0, 1] \cdot d_{t}$ has an opposite sign than inventory $X_t$, $R_{t+1}^{dir}$ is negative. This can be summarised as follows:

\begin{equation}
\begin{aligned}
    \text{Mark-to-Market Value} \quad & M_t = C_t + X_{t} p_t^m \\
                                      & \Delta M_t = \Delta C_{t} + X_{t-1} \Delta p_t^m + \Delta x_t p_t^m \\ %
    \text{PnL Reward} \quad & R_{t+1}^{pnl} = \ln(M_{t}) - \ln(M_{t-1}) \\
    \text{Directional Reward} \quad & R_{t+1}^{dir} = \kappa [-1, 0, 1] \cdot d_{t}  X_{t} \\  %
    \text{Total Reward} \quad & r_{t+1} = w^{dir} R_{t+1}^{dir} + (1-w^{dir})  R_{t+1}^{pnl} \\
\end{aligned}
\end{equation}

The weight on the directional reward $w^{dir} \in [0,1)$ is reduced every learning step by a factor $\psi \in (0,1)$,

$$
w^{dir} \leftarrow \psi w^{dir}
$$
so that initially the agent quickly learns to trade in the signal direction. Over the course of the learning process, $R_t^{pnl}$ becomes dominant and the agent maximises its mark-to-market profits.

\section{Experimental Results} \label{s:results}

We train all RL policies using the problem setup discussed in section \ref{subsec:rl_spec} on 4.5 months of Apple (AAPL) limit order book data (2012-01-01 to 2012-05-16) and evaluate performance on 1.5 months of out-of-sample data (2012-05-17 to 2012-06-31). We only use the first hour of every trading day (09:30 to 10:30) as the opening hour exhibits higher-than-average trading volume and price moves. Each hour of the data corresponds to a single RL episode.
After analysing the results, we also performed a robustness check by repeating the training and analysis pipeline on more recent AAPL data from 2022. Results are reported in section \ref{sec:robustness} and confirm the main conclusions based on earlier data.

Our neural network architecture consists of 3 feed-forward layers, followed by an LSTM layer, for both the value- and advantage stream of the duelling architecture. The LSTM layer allows the agent to efficiently learn a memory-based policy with observations including 100 LOB states.

We compare the resulting learned RL policies to a baseline trading algorithm, which receives the same artificially perturbed high-frequency signal of future mid-prices.
The baseline policy trades aggressively by crossing the spread whenever the signal indicates a directional price move up or down until the inventory constraint is reached. The signal direction in the baseline algorithm is determined as the prediction class with the highest score (down, neutral, or up). When the signal changes from up or down to neutral, indicating no immediate expected price move, the baseline strategy places a passive order to slowly reduce position size until the inventory is cleared. This heuristic utilises the same action space as the RL agent and yielded better performance than trading using only passive orders (placed at the near touch), or only aggressive orders (at the far touch).

\begin{figure*}[h!]
\centering
\includegraphics[width=1\textwidth]{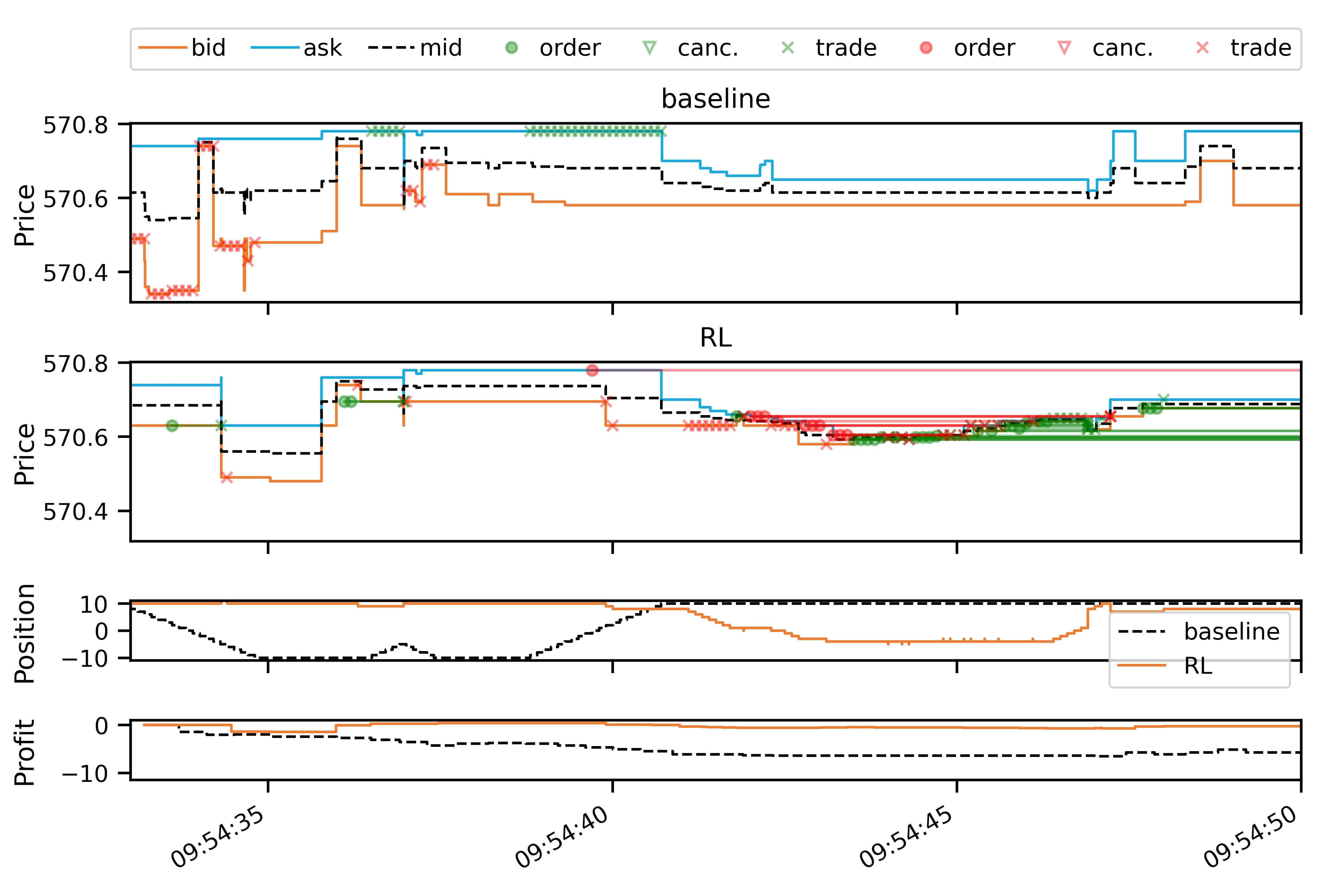}
\caption{A short snapshot of simulation results (AAPL on 2012-06-14), comparing the RL policy (second panel) with the baseline (first panel). The first two panels plot the best bid, ask, and mid-price, overlaying trading events of buy orders (green) and sell orders (red). Circles mark new unmarketable limit orders entering the book. Crosses mark order executions (trades) and triangles order cancellations. Open orders are connected by lines to either cancellations or trades. Since we are simulating the entire LOB, trading activity can be seen to affect bid and ask prices. The third panel plots the evolution of the inventory position of both strategies, and the last panel the trading profits over the period in USD.}
\label{fig:lob_comp}
\end{figure*}

Figure \ref{fig:lob_comp} plots a 17 second simulation window from the test period, comparing the simulated baseline strategy with the RL strategy. It can be seen that prices in the LOB are affected by the trading activity as both strategies inject new order flow into the market, in addition to the historical orders, thereby consuming or adding liquidity at the best bid and ask. During the plotted period, the baseline strategy incurs small losses due to the signal switching between predicting decreasing and increasing future prices. This causes the baseline strategy to trade aggressively, paying the spread with every trade. The RL strategy, on the other hand, navigates this difficult period better by trading more passively out of its long position, and again when building up a new position. Especially in the second half of the depicted time period, the RL strategy adds a large number of passive buy orders (green circles in the second panel of figure \ref{fig:lob_comp}). This is shown by the green straight lines, which connect the orders to their execution or cancellation, some of which occur after the depicted period.

\subsection{Oracle Signal}

The RL agent receives a noisy oracle signal of the mean return $h=10$ seconds into the future (see equation \ref{equ:oracle_signal}). It chooses an action every 0.1s, allowing a sufficiently quick build-up of long or short positions using repeated limit orders of single stocks. The algorithm is constrained to keep the stock inventory within bounds of $[pos_{min}, pos_{max}]=[-10,10]$. To change the amount of noise in the signal, we vary the $a^H$ parameter of the Dirichlet distribution, keeping $a^L=1$ constant in all cases. To keep the notation simple, we hence drop the $H$ superscript and refer to the variable Dirichlet parameter $a^H$ simply as $a$.
We consider three different noise levels, parametrising the Dirichlet distribution with $a=1.6$ (low noise), $a=1.3$ (mid noise), and $a=1.1$ (high noise). A fixed return classification threshold $k=4\cdot 10^{-5}$ was chosen to achieve good performance of the baseline algorithm, placing around 85\% of observations in the up or down category. The signal process persistence parameter is set to $\phi = 0.9$.

\begin{figure*}[h!]
\centering
\includegraphics[width=1\textwidth]{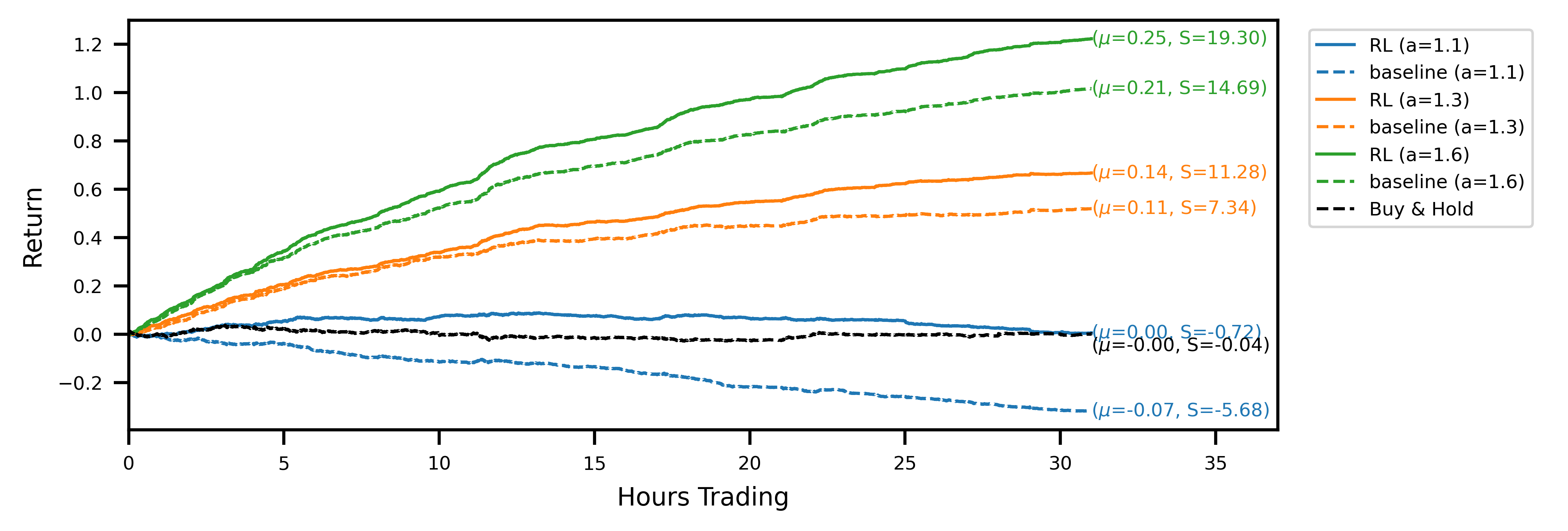}
\caption{Account curves, trading the noisy oracle signal in the test set, comparing the learned RL policies (solid lines) with the baseline trading strategy (dashed). The black line shows the performance of the buy \& hold strategy over the same period. Different colours correspond to different signal noise levels. The RL policy is able to improve the trading performance across all signal noise levels.}
\label{fig:acc_curves}
\end{figure*}

Out-of-sample trading performance is visualised by the account curves in figure \ref{fig:acc_curves}. The curves show the evolution of the portfolio value for a chronological evaluation of all test episodes. Every account curve shows the mean episodic log-return $\mu$ and corresponding Sharpe ratio $S$ next to it. We show that all RL-derived policies are able to outperform their respective baseline strategies for the three noise levels investigated. Over the 31 test episodes, the cumulative RL algorithm out-performance over the baseline strategy ranges between 14.8 ($a=1.3$) and 32.2 ($a=1.1$) percentage points (and 20.7 for $a=1.6$). In the case of the signal with the lowest signal-to-noise ratio (a=1.1), for which the baseline strategy incurs a loss for the test period, the RL agent has learned a trading strategy with an approximately zero mean return. Temporarily, the strategy even produces positive gains. Overall, it produces a sufficiently strong performance to not lose money while still trading actively and incurring transaction costs. Compared to a buy-and-hold strategy over the same time period, the noisy RL strategy similarly produces temporary out-performance, with both account curves ending up flat with a return around zero.
Inspecting Sharpe ratios, we find that using RL to optimise the trading strategy is able to increase Sharpe ratios significantly. The increase in returns of the RL strategies is hence not simply explained by taking on more market risk.

\begin{figure}
     \centering
     \begin{subfigure}[T]{0.49\textwidth}
         \centering
         \includegraphics[width=\textwidth]{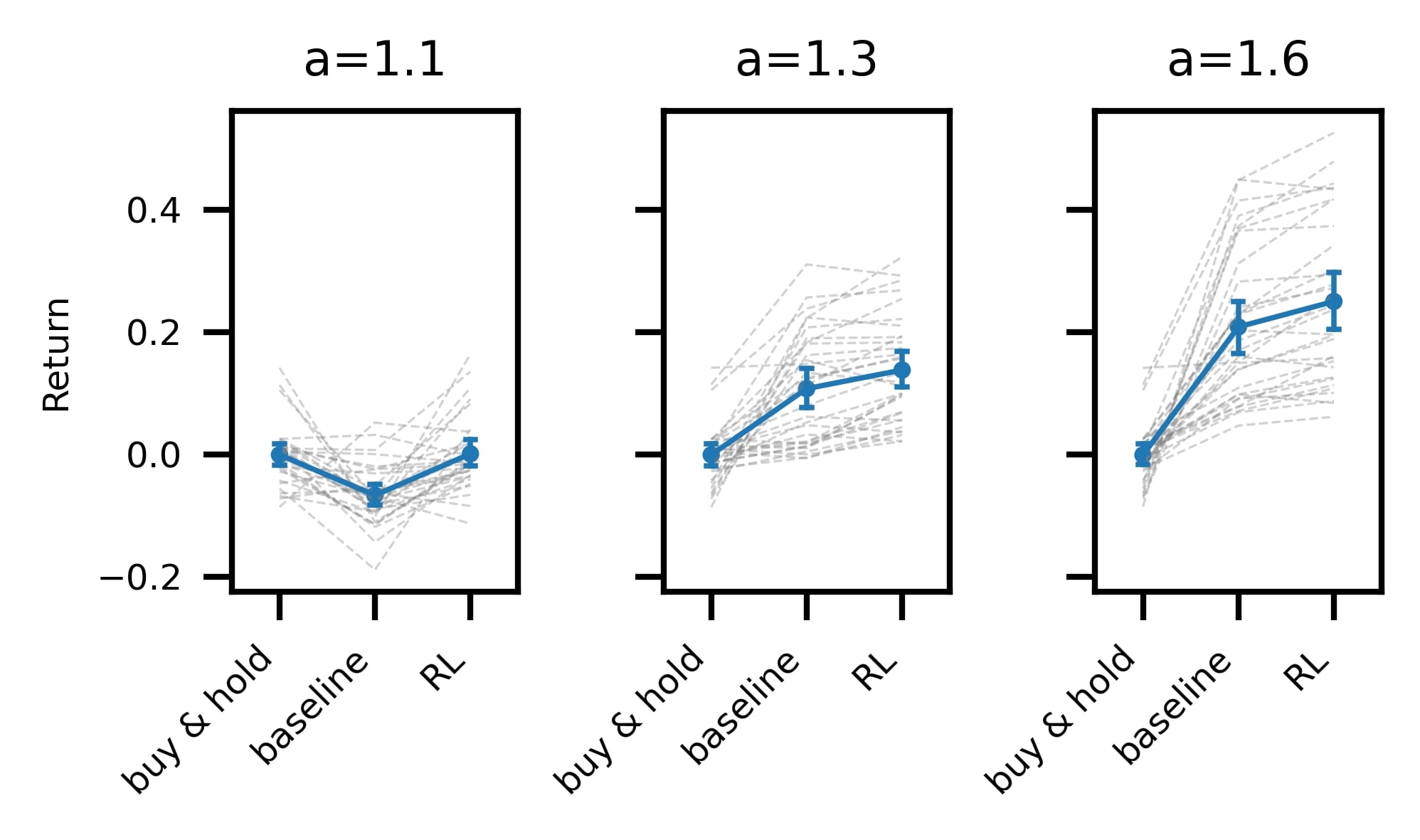}
         \caption{Episodic mean strategy return of buy \& hold, baseline, and RL strategies for high ($a=1.1$), mid ($a=1.3$), and low noise ($a=1.6$) in 31 evaluation episodes. The grey dashed lines connect mean log returns across strategies for all individual episodes. The blue line connects the mean of all episodes with 95\% bootstrapped confidence intervals.}
         \label{fig:mean_comp}
     \end{subfigure}
     \hfill
     \begin{subfigure}[T]{0.49\textwidth}
         \centering
         \includegraphics[width=\textwidth]{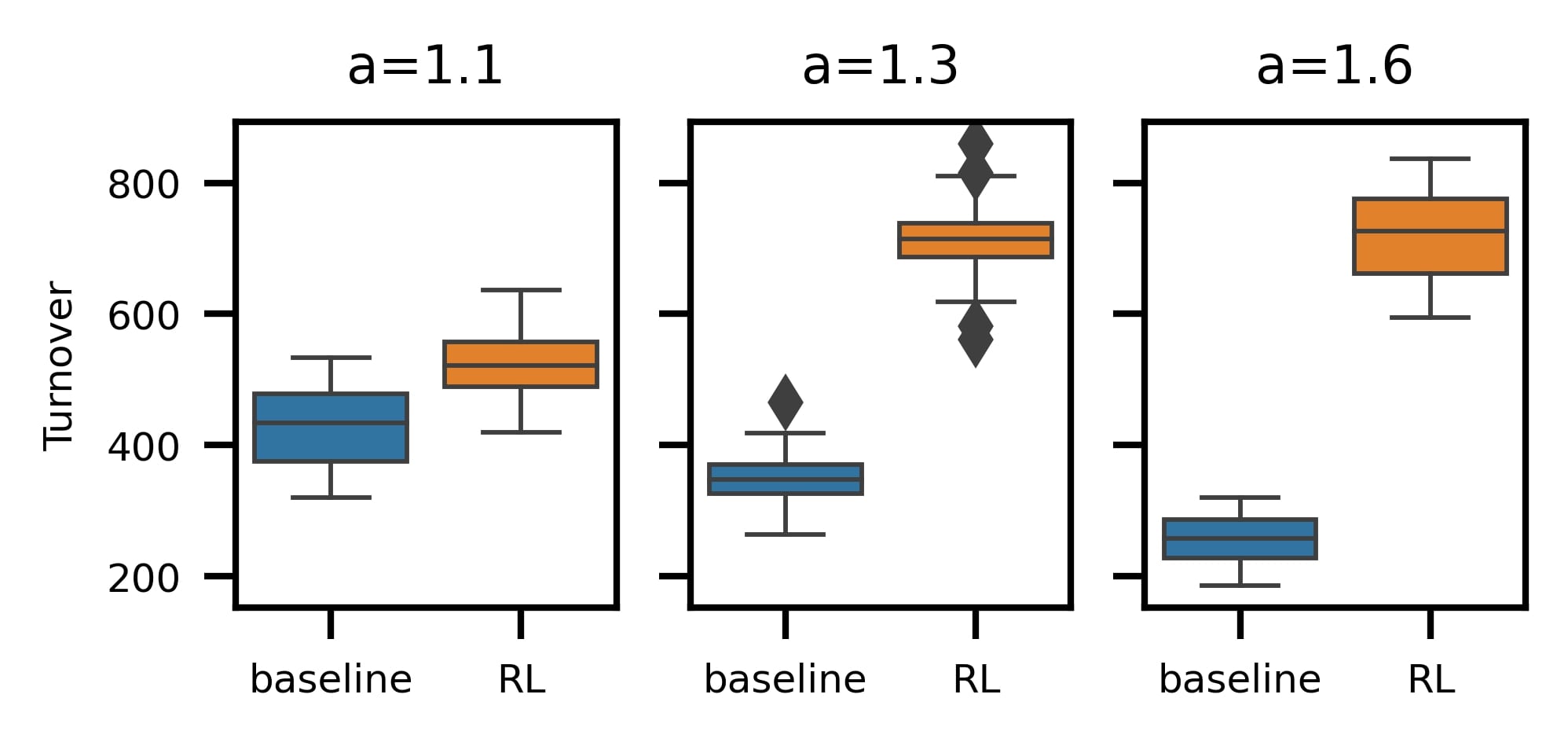}
         \caption{Turnover per episode: comparison between baseline and RL strategy. Lower noise results in a more persistent signal, decreasing baseline turnover, but a higher quality signal, resulting in the RL policy to increase trading activity and turnover.}
         \label{fig:turnover}
     \end{subfigure}
        \caption{Mean return and turnover of the baseline and RL trading strategies.}
        \label{fig: subfigures}
\end{figure}

Figure \ref{fig:mean_comp} compares the mean return between the buy \& hold, baseline, and RL policies for all out-of-sample episodes across the three noise levels. 
A single dashed grey line connects the return for a single test episode across the three trading strategies: buy \& hold, baseline, and the RL policy. The solid blue lines representing the mean return across all episodes. Error bars represent the 95\% bootstrapped confidence intervals for the means. 
Testing for the significance of the differences between RL and baseline returns across all episodes (t-test) is statistically significant ($p\ll 0.1$) for all noise levels. Differences in Sharpe ratios are similarly significant. We can thus conclude that the high frequency trading strategies learned by RL outperform our baseline strategy for all levels of noise we have considered.

It is also informative to compare the amount of trading activity between the baseline and RL strategies (see figure \ref{fig:turnover}). The baseline turnover decreases with an increasing signal-to-noise ratio (higher $a$), as the signal remains more stable over time, resulting in fewer trades. In contrast, the turnover of the RL trading agent increases with a higher signal-to-noise ratio, suggesting that the agent learns to trust the signal more and reflecting that higher transaction costs, resulting from the higher trading activity, can be sustained, given a higher quality signal. In the high noise case ($a=1.1$), the RL agent learns to reduce trading activity relative to the other RL strategies, thereby essentially filtering the signal. The turnover is high in all cases due to the high frequency of the signal and the fact that we are only trading a small inventory. Nonetheless, performance is calculated net of spread-based transaction costs as our simulator adequately accounts for the execution of individual orders.

\begin{table}[h!]
\centering
\begin{tabular}{@{}llll@{}}
\toprule
                                     & a=1.1   & a=1.3   & a=1.6   \\ \midrule
action skipped [\%]                  & 24.5       & 43.8       & 7.8       \\
sell levels (bid, mid, ask) {[}\%{]} & (95.4, 3.1, 1.5) & (94.6, 2.8, 2.65) & (97.2, 1.9, 0.9) \\
buy levels (bid, mid, ask) {[}\%{]}  & (1.1, 1.3, 97.5) & (1.6, 52.9, 45.5) & (1.7, 13.0, 85.3) \\ \bottomrule
\end{tabular}%
\caption{Actions taken by RL policy for the three different noise levels: the first row shows how often the policy chooses the ``skip'' action. Not choosing this action does however not necessarily result in an order being placed, e.g. if inventory constraints are binding. The last two rows show the relative proportion of limit order placement levels for sell orders, and buy orders, respectively.}
\label{tab:actions}
\end{table}

Table \ref{tab:actions} lists action statistics for all RL policies, including how often actions are skipped, and the price levels at which limit orders are placed, grouped by buy and sell orders. With the least informative signal, the strategy almost exclusively uses marketable limit orders, with buy orders being placed at the bid and sell orders at the ask price. With better signals being available ($a=1.3$ and $a=1.6$), buy orders are more often placed at the mid-quote price, thereby trading less aggressively and saving on transaction costs. Overall, the strategies trained on different signals all place the majority of sell orders at the best bid price, with the amount of skipped actions varying considerably across the signals.

\subsection{Robustness Evaluation on Recent Data} \label{sec:robustness}

\begin{figure*}[h!]
\centering
\includegraphics[width=0.45\textwidth]{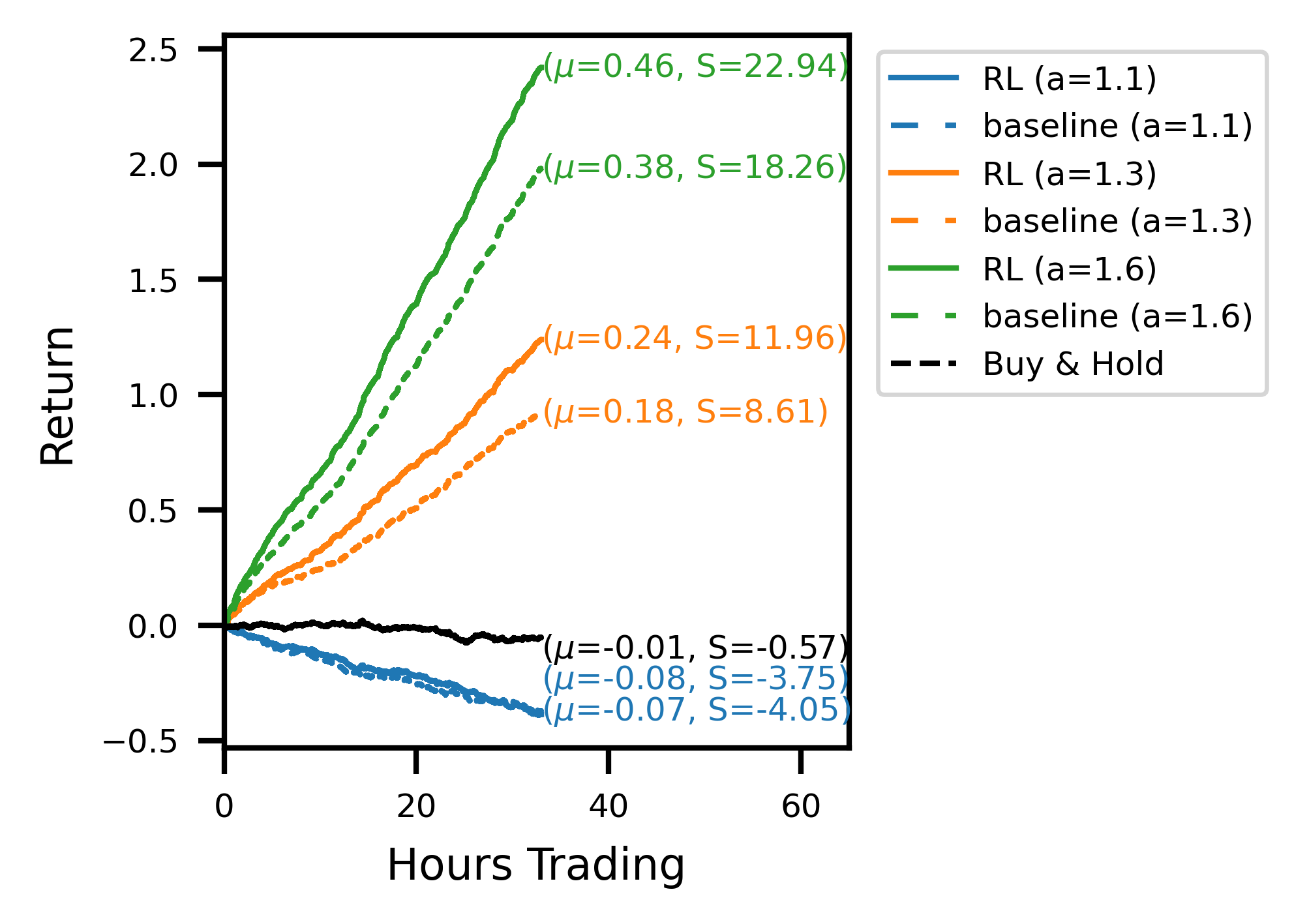}
\caption{Robustness check: account curves for second evaluation period 2022-11-14 to 2022-12-31, comparing RL performance to baselines for three noise levels. Across all noise levels, the RL strategies produce a similar level of out-performance during this period than during the original period in 2012. For a high noise signal ($a=1.1$), neither the baseline nor the RL policies are profitable. In this case, our RL strategy performs similarly to the baseline and does not learn to stop trading completely.
This supports the hypothesis that similar micro-structure effects can be utilised in more recent periods, during a different macroeconomic landscape.}
\label{fig:robustness}
\end{figure*}

To evaluate our results on more recent LOB data, we also trained RL policies using 4.5 months of AAPL LOB data from the second half of 2022 (2022-07-01 to 2022-11-13) and evaluated on 1.5 months of held-out test data (2022-11-14 to 2022-12-31). Figure \ref{fig:robustness} shows account curves over the evaluation period for all 3 noise levels. For low ($a=1.6$) and medium ($a=1.3$) noise levels, the RL policies, again, beat the baselines by a significant margin and increase profits significantly. For a low-quality signal with high noise levels ($a=1.1$), the RL policy performs similarly to the losing baseline strategy and also doesn't make a profit. Interestingly, the strategy also does not learn to stop trading altogether, which would represent a superior policy in this case. This could be due to a common problem with RL: effective exploration. The local optimum of not trading could not be found using an epsilon-greedy exploration policy in this case.
Overall, even though macroeconomic and financial market conditions in 2022 differed markedly from 2012, our results support the conclusion that the RL policies can utilise similar market micro-structure effects in both periods to improve the execution of a trading strategy based on a price forecast signal.

\section{Conclusions} \label{s:conclusion}

Using Deep Double Duelling Q-learning with asynchronous experience replay, a state-of-the-art off-policy reinforcement learning algorithm, we train a limit order trading strategy in an environment using historic market-by-order (MBO) exchange message data. For this purpose we develop an RL environment based on the ABIDES \cite{abides_byrd20} market simulator, which reconstructs order book states dynamically from MBO data. Observing an artificial high-frequency signal of the mean return over the following 10 seconds, the RL policy successfully transforms a directional signal into a limit order trading strategy. The policies acquired by RL outperform our baseline trading algorithm, which places marketable limit orders to trade into positions and passive limit orders to exit positions, both in terms of mean return and Sharpe ratio. We investigate the effect of different levels of noise in the alpha signal on the RL performance. Unsurprisingly, more accurate signals lead to higher trading returns but we also find that RL provides a similar added benefit to trading performance across all noise levels investigated. 

The task of converting high-frequency forecasts into tradeable and profitable strategies is difficult to solve as transaction costs, due to high portfolio turnover, can have a prohibitively large impact on the bottom line profits. We suggest that RL can be a useful tool to perform this translational role and learn optimal strategies for a specific signal and market combination. We have shown that tailoring strategies in this way can significantly improve performance, and eliminates the need for manually fine-tuning execution strategies for different markets and signals. For practical applications, multiple different signals could even be combined into a single observation space. That way the problem of integrating different forecasts into a single coherent trading strategy could be directly integrated into the RL problem.

A difficulty for all data-driven simulations of trading strategies relying on market micro-structure is accurately estimating market impact. We address this partially by injecting new orders into historical order streams, thereby removing liquidity from the LOB. If liquidity at the best price is used up, this would automatically increase transaction costs by consuming deeper levels of the book. This mechanism accurately models temporary \emph{direct} market impact, but cannot take account of the \emph{indirect} or permanent component due to other market participants' reactions to our orders. By only allowing to trade a single stock each time step, we posed the problem in a way to minimise the potential effect indirect market impact would have on the performance of the RL strategy. The strategy trades small quantities on both sides of the book without accumulating large inventories. Such trading strategies are \emph{capacity} constrained by the volume available in the book at any time and belong to a different class than \emph{impact} constrained strategies, which build up large inventories by successively submitting orders in only one direction.
Furthermore, we measure trading performance relative to a baseline strategy, which makes the same assumptions on market impact. However, accurately modelling the full market impact of high-frequency trading in LOB markets in a data-driven approach is an interesting direction for future research and would allow evaluating strategies with larger order sizes. Recent attempts in this vein have used agent-based models \cite{byrd2020abides} or generative models \cite{coletta2021towards, coletta2022learning, coletta2023conditional}.

We chose to focus our investigations in this paper on AAPL stock as a challenging test case of a small-tick stock, i.e. one where the minimum tick size is small relative to the stock price, with a high trading volume. Showing that we can train an RL agent to improve the profitability of an alpha signal in this example, indicates that similar performance improvements could be possible in larger-tick stocks with less trading activity.
Although results are limited to a single company due to computational constraints, functional relationships in the micro-structure of the market have been found to be stable over time and across companies in prior work.
In a large-scale study of order flow in US equity LOBs \citet{sirignano2019universal} found a \emph{universal} and \emph{stationary} relationship between order flow and price changes, driven by robust underlying supply and demand dynamics. 
Similarly, supervised training of deep neural networks to predict the mid-price direction a few ticks into the future has been shown to work for a wide range of stocks \cite{deeplob_zhang19}. In contrast to lower-frequency trading strategies, whose performance often varies with market conditions, such as the presence of price trends or macroeconomic conditions, high-frequency strategies do not suffer the same degree of variability. Nonetheless, a systematic investigation of potential changes in LOB dynamics due to crisis periods or rare events could be an interesting avenue for future research.

While we here show an interesting use case of RL in limit order book markets, we also want to motivate the need for further research in this area. There are many years of high-frequency market data available, which ought to be utilised to make further progress in LOB-based tasks and improve RL in noisy environments. 
This, together with the newest type of neural network architectures, such as attention-based transformers \cite{attention_vaswani17, sparse_transformers_child19}, enables learning tasks in LOB environments directly from raw data with even better performance.
For the task we have considered in this paper, future research could enlarge the action space, allowing for the placement of limit orders deeper into the book and larger order sizes. Allowing for larger sizes however would require a realistic model of market impact, considering the reaction of other market participants.

\clearpage

\bibliographystyle{unsrtnat}  
\bibliography{references}  %

\clearpage
\section{Appendix}

We use the RLlib library \cite{rllib_liang18} for a reference implementation of the APEX algorithm. Table \ref{tab:params} shows a selection of relevant parameters we used for RL training.

\begin{table}[h] \centering
\begin{tabular}{ll}
\hline
Paramter                      & Value                               \\ \hline
timesteps\_total              & 300e6                               \\
framework                     & torch                               \\
num\_gpus                     & 1                                   \\
num\_workers                  & 42                                  \\
batch\_mode                   & truncate\_episode                   \\
gamma                         & .99                                 \\
lr\_schedule                  & {[}{[}0,2e-5{]}, {[}1e6, 5e-6{]}{]} \\
buffer\_size                  & 2e6                                 \\
learning\_starts              & 5000                                \\
train\_batch\_size            & 50                                  \\
rollout\_fragment\_length     & 50                                  \\
target\_network\_update\_freq & 5000                                \\
n\_step                       & 3                                   \\
prioritized\_replay           & False                              
\end{tabular}
\caption{Selected RL parameters for APEX algorithm using RLlib \cite{rllib_liang18} library for training.}
\label{tab:params}
\end{table}

\begin{figure}[h]
\centering
\includegraphics[width=0.75\columnwidth]{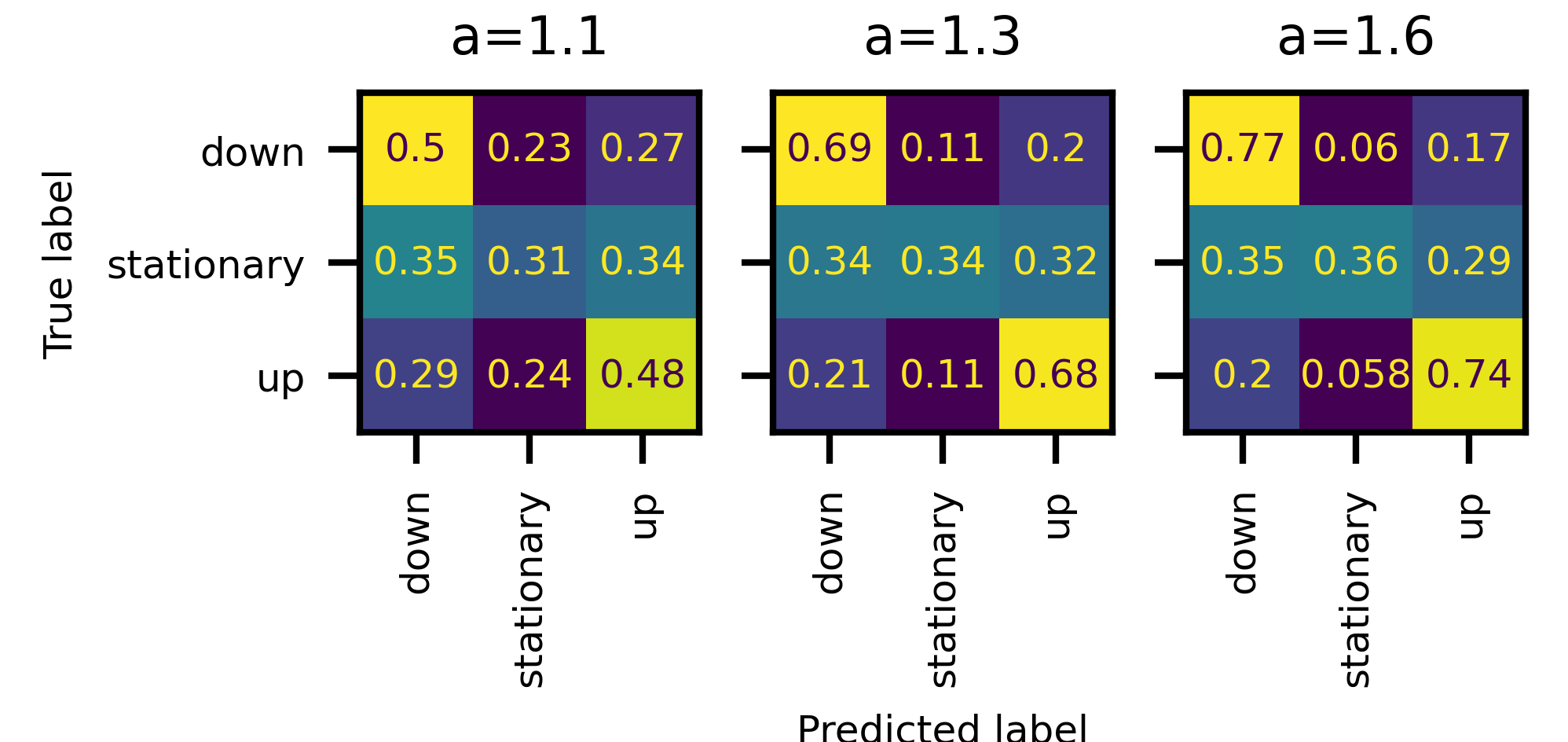}
\caption{Confusion matrices of the artificial oracle signal for three noise levels, from low to high noise.}
\label{fig:conf_matrices}
\end{figure}

Figure \ref{fig:conf_matrices} shows confusion matrices interpreting the oracle signal scores as probabilities over the three classes: down, stationary, and up. The predicted class is thus the one with the highest score.

\end{document}